\documentclass[transmag]{IEEEtran}
\usepackage{latexsym}
\usepackage{epsfig}
\usepackage{amsfonts,amsmath,amssymb,amstext}
\usepackage{epstopdf}
\usepackage{graphicx}
\usepackage{psfrag}
\usepackage{dsfont}
\usepackage{color}
\usepackage{cite}
\usepackage{ngerman}
\usepackage{flushend}
\usepackage{mathrsfs}
\def\BibTeX{{\rm B\kern-.05em{\sc i\kern-.025em b}\kern-.08em T\kern-.1667em\lower.7ex\hbox{E}\kern-.125emX}}

\begin{document}
\title{Reconfigurable Intelligent Surfaces for Smart Cities:\\ Research Challenges and Opportunities}
\author{Steven Kisseleff,~\IEEEmembership{Member,~IEEE,} Wallace A. Martins,~\IEEEmembership{Senior Member,~IEEE,}\\Hayder Al-Hraishawi,~\IEEEmembership{Member,~IEEE,} Symeon Chatzinotas,~\IEEEmembership{Senior Member,~IEEE,} Bj\"orn Ottersten,~\IEEEmembership{Fellow,~IEEE}
\thanks{
This work was supported by Luxembourg National Research Fund (FNR) under the CORE projects 5G-Sky and CI-PHY.}
\thanks{
The authors are with the Interdisciplinary Centre for Security, Reliability and Trust (SnT), University of Luxembourg, Luxembourg. \textit{Corresponding author: Steven Kisseleff (steven.kisseleff@uni.lu).}
}}
\IEEEtitleabstractindextext{
\begin{abstract}
The concept of Smart Cities has been introduced as a way to benefit from the digitization of various ecosystems at a city level. To support this concept, future communication networks need to be carefully designed with respect to the city infrastructure and utilization of resources. Recently, the idea of `smart' environment, which takes advantage of the infrastructure for better performance of wireless networks, has been proposed. This idea is aligned with the recent advances in design of reconfigurable intelligent surfaces (RISs), which are planar structures with the capability to reflect impinging electromagnetic waves toward preferred directions. Thus, RISs are expected to provide the necessary flexibility for the design of the `smart' communication environment, which can be optimally shaped to enable cost- and energy-efficient signal transmissions where needed. Upon deployment of RISs, the ecosystem of the Smart Cities would become even more controllable and adaptable, which would subsequently ease the implementation of future communication networks in urban areas and boost the interconnection among private households and public services. In this paper, we describe our vision of the application of RISs in future Smart Cities. In particular, the research challenges and opportunities are addressed. The contribution paves the road to a systematic design of  RIS-assisted communication networks for Smart Cities in the years to come.
\end{abstract}
}
\maketitle
\section{Introduction}
\label{sec:1}
\IEEEPARstart{R}{econfigurable} intelligent surfaces (RISs) are emerging as a promising technology to meet the ever-increasing demands of wireless networks beyond 5G~\cite{Liaskos}. RISs are intelligently designed artificial planar structures with reconfigurable properties enabled via integrated electronic circuits, which can be programmed to reflect an impinging electromagnetic wave in a controlled manner. RISs are manufactured with low-profile, light weight, cheap materials that can be shaped with conformal geometries, thus easing their deployment on a variety of environment surfaces, such as facades of buildings, walls, ceilings, etc.~\cite{Coquet2018}. The signal propagation from transmitters to receivers can therefore be assisted by steering the RIS-reflected signals in directions that enhance the resulting signal quality, which in turn can be exploited to attain substantially higher spectral efficiencies compared to current wireless systems~\cite{Renzo2019}. Furthermore, co-channel interference can be reduced or avoided by choosing different propagation paths for the interfering signals. This is especially important in urban environments with dense deployment of terminals.


Smart Cities are an emerging concept relying on the harmonization of  digital technologies at a city level~\cite{albino2015smart}. The idea is to improve accessibility to public services, advance digitization of the urban environment, and monitor various societal processes as well as city assets in a large scale. More and more cities worldwide are deploying Smart City-related components at different levels, e.g. with respect to the city administration (e-governance)~\cite{roy2006government, paskaleva2009enabling}. In the foreseeable future, the Smart City components are expected to become a unified and controllable entity with potential features of self-management~\cite{8307785}. Another driver of the digitization and Smart City evolution is the deployment of the Internet of things (IoT) with a large number of monitoring devices connected to the city infrastructure. These devices will provide advanced monitoring capabilities, from which new services of Smart Cities will emerge~\cite{6740844}.

\begin{figure}
    \centering
    \includegraphics[width=0.49\textwidth]{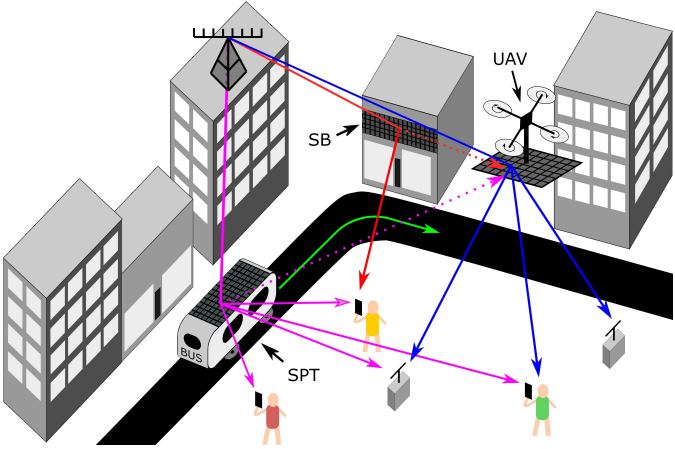}
    \caption{Example of RIS-assisted signal propagation in Smart Cities, where Smart Buildings (SBs), unmanned aerial vehicles (UAVs) and even smart public transport vehicles (SPTs) are equipped with RIS and collaborate to enhance the connectivity and functionality of Smart Cities. Signals reflected from SB, UAV and SPT are indicated by red, blue and purple arrows. The trajectory for the motion of the bus is depicted with a green arrow.}
    \label{fig:SC_RIS}
\end{figure}

The seamless integration of future communication networks with the Smart City components is necessary for a successful digital transformation. In this context, RIS might be the ideal candidate technology to comprise the basic infrastructure of those future networks, since the RIS-assisted communication environment of a city  would itself become a controllable asset (`smart environment’~\cite{Renzo2019}). Thus, a symbiotic relation between Smart Cities and their communication infrastructures is envisioned. On the one hand, future communication systems would exploit the benefits of the controllable Smart City environment, such as improved quality-of-service (QoS), resource utilization, and security. On the other hand, Smart Cities would certainly benefit from a flexible and context-oriented broadband connectivity between the Smart City public service providers, private households/entities and sensor networks. For this, RIS should be deployed at as many objects as possible, including both stationary (e.g. smart buildings) and mobile (e.g. vehicles) objects, so that the connectivity among base stations (BSs), RISs, and end-users/terminals would be improved; see Fig.~\ref{fig:SC_RIS} for an illustration. Through this, the Smart City's inherent intelligence can be further nourished by the added intelligence of the supporting communication infrastructure.

A distinct advantage of employing RISs in Smart Cities is the substantially reduced energy consumption for powering wireless systems, given the expected energy efficiency gains (by an order of magnitude) associated with RIS~\cite{Huang2019}. Accordingly, the deployment of RISs in future Smart Cities will help  mitigate the ecological concerns raised in various countries with respect to the utilization of fossil fuel. Furthermore, the controllable signal propagation can relieve human exposure to RF radiations, thus addressing some health concerns that are raised due to the proliferation of wireless technologies and devices, especially in very high frequencies, such as millimeter-wave and THz communications. 

In order to understand how the intelligence of RIS can be incorporated into the Smart City environment, we provide a review of existing works in the following.
\subsection{Prior works}
\subsubsection{Review of RIS}
The performance of a wireless communication system is ultimately determined by the underlying communication channel, which is  traditionally considered to be an  uncontrollable entity, being usually characterized by some probabilistic model. In this context, most of the conventional wireless communication techniques (e.g., beamforming and channel coding) either neutralize or exploit the propagation effects between  transmitting and receiving ends, starting from the assumption that the channel behavior cannot be controlled. 

RIS is a disruptive concept that can be used for enhancing the performance of communication systems by controlling the propagation of travelling electromagnetic waves, owing to the added capability of real-time reconfigurability and control of the RIS's adjustable antenna elements. 

RIS has been named a new paradigm in wireless communications in~\cite{Liaskos}. The authors proposed RIS as one of the key enablers for the 5G and beyond systems. In particular, research challenges and directions for the design and deployment of RIS based on the prior work~\cite{Yang} have been addressed. Since RIS represents a fundamentally new concept of signal propagation, channel modeling and acquisition are of utmost importance for the system design. Hence, multiple works have been dedicated to these topics, e.g.~\cite{Zheng2019,jensen2019}. Furthermore, advanced techniques of multi-antenna transmissions need to be developed taking into account the peculiarities associated with controllable signal reflections. An additional motivation for the theoretical investigations in this area has been provided in~\cite{Cui,Tan}, where the first experimental results for RIS-assisted signal propagation have been presented.  

In parallel to the mainstream investigations on RIS, a similar technology, named large intelligent surfaces (LIS), has been proposed in~\cite{Hu, Huang}, wherein the surfaces can be operated in one of the two modes, namely: passive and active modes. LIS resemble the functionality of RIS in passive mode, whereas they act as a massive antenna array in active mode. Based on these works, joint active and passive beamforming using RIS have been proposed for multiuser (MU) networks in~\cite{Huang2018a, DiJSAC2020, GuoTWC2020}. In these works, the main aspects of resource allocation have been addressed as well. 

Most of the relevant works in the research field of RIS focus on indoor scenarios~\cite{Liaskos,Yang,Cui,Tan, Hu,Huang,Huang2018a,GuoTWC2020,DiJSAC2020}, where the reflections from the walls can be potentially very harmful without RIS. Using RIS, the number of connected devices in such scenarios can be increased without causing a performance degradation for the network due to harmful co-channel interference signals. Outdoors, the target application would be Smart City. So far, however, only a few works have investigated RIS-assisted outdoor scenarios~\cite{Ma, Li}. The main innovation addressed in these works is the use of unmanned aerial vehicles (UAVs). In such scenarios, UAVs can act not only as receivers, but also as mobile relays to serve as access points for multiple terminals. In particular, the beamforming toward UAVs and the tracking of the UAV position by the BS has been investigated in~\cite{Li}.

Recently, several surveys on the advances in RIS technology and the creation of a smart environment have been published. For further details on RIS, we refer the reader to~\cite{Gong,di2020smart}, where all relevant studies have been described and future challenges in a general context have been addressed. 
\subsubsection{Review of Smart Cities}
The initiative of moving toward Smart Cities is closely related to the digital transformation~\cite{tomivcic2019smart, cocchia2014smart}. Here, the triggering factor for the Smart Cities is the ongoing urbanization and growing population~\cite{caragliu2011smart}. Novel technologies offer cities the possibility to contribute to the well-being of the citizens by increasing the number and variety of services while reducing the long-term operation costs. Furthermore, the decision makers (mayors, policy makers, and business developers) are expected to rely more and more on the situational awareness acquired by real-time monitoring of processes and assets in the cities, so that those players can both establish a regulatory framework as well as take informed decisions that are capable of quickly reacting, or even anticipating, to  emergencies~\cite{sarker2018smart, 8307785}. 

The Smart City concept is closely related to the older concepts of Smart Home, Smart Factory (Industry 4.0), Smart Grid, etc. All of these applications appeared independently and have seen promising developments in the recent years. Nevertheless, a generalization and unification of these independent components is envisioned and, according to the increasing interest by the research community, will be addressed very soon as well.

Urban challenges associated with Smart Cities can be tackled using the broadband connectivity, a wide range of low-cost real-time sensors, machine-to-machine communication, cloudification and virtualization of services, big data analytics, and visualization tools~\cite{brandt2018smart, BIBRI2018230, 6740844}. Hence, the envisioned digital transformation is empowered by the technological progress in the field of wireless communications and computer science. 

For a better insight into the Smart City technology, let us consider the basic structure of the Smart City model with the following components~\cite{gaur2015smart}:
\begin{itemize}
    \item Services/Applications
    \item Platforms
    \item Communication 
    \item Sensors/Actuators
\end{itemize}
These components represent architectural layers, which need to be carefully designed in accordance with the service requirements. The services and applications are dictated by the respective ecosystems connected to the Smart City. The platforms are provided by the stakeholders and should move toward open access and non-proprietary platforms according to the current trend. The layer of sensors and actuators will be based upon the growing infrastructure of the IoT, which imposes massive machine-type connectivity to be provided by the Smart City. Besides IoT, traditional user equipment, such as smartphones and notebooks, can be connected to the Smart City via specific interfaces (if available). Nevertheless, the connection between the upper layers, i.e. platforms and services, and lower layer, i.e. IoT, is established using wireless communications. Hence, wireless technology plays a fundamental role as an enabler for Smart Cities. 

All these layers are vulnerable to security and privacy attacks, such that sufficient security is required throughout the whole architecture. This problem is even more challenging given the distributed nature and scalability of the resulting system~\cite{eckhoff2017privacy, zhang2017security}. On the other hand, the objective to provide massive/universal connectivity is very challenging as well. In order to guarantee both the QoS and the target communication security, it is of paramount importance to substantially improve the performance of the main system enabler, namely the underlying communication system.

There are many works describing various visions for Smart Cities and corresponding enabling technologies. For more information on this topic, we refer the reader to~\cite{8307785}.

\subsection{Contributions}
In this work, we focus on the research challenges for the development of innovative communication concepts based on RIS within the scope of Smart Cities. For this, we consider the promising configurations of RIS-assisted networks and discuss the most likely design problems and re-architecturing directions to integrate RISs into Smart City infrastructures in order to harness their properties of scalability, spectral efficiency, resiliency, and energy efficiency. 
Hence, the main contributions of this paper are summarized as follows:
\begin{itemize}
    \item A brief technical description is provided on the existing RIS technology and on the design of RIS-assisted communications.
    \item The key use cases for RIS deployment in Smart Cities are identified and their peculiarities with respect to RIS-assisted signal propagation are discussed.
    \item The research challenges pertained to the design of wireless communication networks, which need to be faced in the near future, are identified. Some examples are formulated mathematically, which should substantially facilitate future research, since each problem can be readily addressed based on our work.
    \item Further research challenges and opportunities are explored, which are likely to be targeted in the long run.
\end{itemize}
Throughout the open literature, the research challenges of integrating RISs into wireless networks are solely discussed within the existing network architectures and the traditional urban ecology~\cite{Gong}. However, today’s landscape of cities and the corresponding communication infrastructure will soon be supplanted by new generations of wireless systems to complement the digital transformation toward Smart Cities. This observation motivates us to fill this gap in the existing literature by providing a vision of RIS-assisted Smart City networks along with the proposal of some relevant research challenges.
Specifically, the contributions of this work provide a high educational and technical value for researchers and system designers, since the potential research challenges and scenarios are elaborated and their importance is clarified. Hence, it is possible to prioritize challenges that need to be assessed in the near future and consider possible directions/opportunities in the far future.  

This paper is organized as follows. Section~\ref{sec:2} provides an overview of the existing state-of-the-art methods and system configurations of RIS-assisted communications. This includes signal processing a t transmitters and  receivers as well as the general end-to-end system model. The role of RISs within the Smart City concept is explained in Section~\ref{sec:3}, in which various scenarios/use cases are addressed. In addition, the differences with respect to signal propagation and likely design approaches are pointed out. The existing research challenges for the design of large RIS-assisted networks in Smart Cities, expected to be addressed in the near future,  are discussed in Section~\ref{sec:4}. Further research challenges and opportunities for the more distant future are described in Section~\ref{sec:5}. The paper is then concluded in Section~\ref{sec:6}. 

\subsubsection*{Notation}
Scalars are denoted by italic letters, while vectors and matrices 
are denoted by boldface letters (lowercase for vectors and uppercase for matrices). 
Calligraphic letters, like $\cal F$, denote sets. The set of complex, real, integer, and natural numbers are respectively denoted by $\mathbb{C}, \mathbb{R}, \mathbb{Z}$, and $\mathbb{N}$. The notations $\Re\left\{\cdot\right\}$ and $\Im\left\{\cdot\right\}$ denote the real and imaginary parts of a complex argument, respectively. The rectifier operator is defined as $\left[x\right]^+ = \max\{0,x\}$, for any $x\in\mathbb{R}$. The notations $(\cdot)^{\rm T}$, $(\cdot)^{\rm H}$, and $(\cdot)^\dagger$ stand for transpose,  Hermitian transpose, and pseudo-inversion  operations on $(\cdot)$, respectively. The Frobenius-norm and 2-norm are respectively denoted by $\Vert\cdot\Vert_{\rm F}$ and $\Vert\cdot\Vert_2$.

\section{Existing RIS Technology} 
\label{sec:2}
This section reviews the existing methods for the design of RIS-assisted wireless networks. It starts with a general system model for the multiuser communication and moves on to the specific aspects of the main system components and signal processing therein.

\subsection{System model}
\begin{figure}
    \centering
    \includegraphics[width=0.49\textwidth]{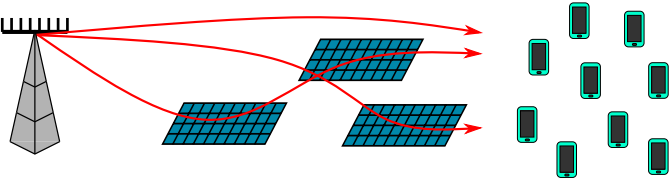}
    \caption{Example of signal propagation in smart environments. Signals from the BS are passively relayed to the terminals.}
    \label{fig:sysmod}
\end{figure}

The system model of RIS-assisted communication networks has a higher diversity of components than usual communication systems. In particular, smart environments can include a large number of RISs and active relays with arbitrary or structured deployment. For the sake of simplicity, we shall initially focus on RIS-assisted environments  without the use of active relays.\footnote{Active relaying, hybrid relaying, and mobile multihop relaying are considered promising research directions in the Smart City context and will be covered in Sections~\ref{sec:4} and~\ref{sec:5}.} Also, the deployment of RISs and the distribution of users depend on the specific scenario, which will be further detailed in Section~\ref{sec:3}. Accordingly, the basic system model described here consists of a BS equipped with multiple transmit/receive antennas, single antenna terminals, and multiple RISs as passive relay devices, as illustrated in Fig.~\ref{fig:sysmod}.
More specifically, a BS equipped with $L$ transmit antennas, $N$ RIS devices, and $K$ single-antenna terminals are  assumed. Clearly, there are many paths for the signal propagation, including a waveguide-like propagation  where the signal is guided  with reduced power losses via consecutive reflections from one RIS to another. The resulting transmission channel may depend non-linearly on the parameters and number of RISs. 

The signal propagation for downlink transmissions\footnote{Although we shall not address uplink transmissions here, they are also of interest for future RIS-assisted communication systems.} can be described using a common matrix notation:
\begin{equation}
    \mathbf{y}[t]=\mathbf{H}\left(\boldsymbol{\theta}^{(1)},\ldots,\boldsymbol{\theta}^{(N)}\right)\![t]\cdot\mathbf{x}[t]+\mathbf{w}[t],
\end{equation}
where $\mathbf{x}[t]\in \mathbb{C}^{L\times 1}$ and $\mathbf{y}[t]\in \mathbb{C}^{K\times 1}$ denote the transmitted and received signal vectors, respectively, during the $t$-th symbol interval, with $t\in\mathbb{Z}$. In addition, $\mathbf{w}[t]$ represents a realization of a zero-mean wide-sense stationary (WSS) additive white Gaussian noise (AWGN) with covariance matrix ${\rm diag}\left\{\sigma_{w,1}^2,\dots,\sigma_{w,K}^2\right\}$, where $\sigma_{w,k}^2$ is the noise power associated with the $k$-th terminal. Note that RISs only reflect the signals without injecting any additional noise signals unlike traditional active relays. Furthermore, the downlink channel matrix $\mathbf{H}\left(\boldsymbol{\theta}^{(1)},\ldots,\boldsymbol{\theta}^{(N)}\right)\![t]\in \mathbb{C}^{K\times L}$ is a function of the parameters $\boldsymbol{\theta}^{(n)}\in\mathbb{C}^{Q^{(n)}\times 1}$, with $n\in\{1, \dots, N\}$, comprising the reflection coefficients associated with the $Q^{(n)}$ reflective elements/planes of the $n$-th RIS.\footnote{One usually has $Q^{(n)} \gg L \geq K$ for all $n\in\{1, \dots, N\}$.} In this work, the notation $(\cdot)
^{(n)}$ will be consistently employed to refer to a mathematical object (scalar, vector, matrix) associated with the $n$-th RIS, for $n\in\{1, \dots, N\}$. 

It is assumed that the $q$-th complex reflection coefficient of the $n$-th RIS can be written as
\begin{align}
    \theta_{q}^{(n)} \!= \eta_{q}^{(n)} \!\cdot{\rm e}^{{\rm j}\phi_{q}^{(n)}}\!,\forall (n,q)\in\{1,\dots,N\}\times\{1,\dots,Q^{(n)}\},
\end{align}
with $\phi_{q}^{(n)} \in [0,2\pi)$ denoting the corresponding phase shift, and $\eta_{q}^{(n)} \in [0,1]$ denoting the corresponding reflection efficiency. An interesting feature of RISs is the possibility to configure the parameters $(\phi_{q}^{(n)}, \eta_{q}^{(n)})$ so that the travelling signals are not simply scattered but guided to an intended receiver. This aspect makes RIS-enabled channels vary with the patterns of the RIS's reflective elements.  In this context, establishing suitable path-loss models that capture all nuances of changes induced by the presence of RISs is essential for the corresponding performance evaluation.

Given that most of the models discussed here apply to a single snapshot, the dependency on the parameter $t$ will be omitted in the following discussions, for the sake of notation simplicity. 

A simple model for the downlink channel matrix is 
\begin{align}\label{eq:HDL}
    \mathbf{H}&\left(\boldsymbol{\theta}^{(1)},\ldots,\boldsymbol{\theta}^{(N)}\right) = \underbrace{\mathbf{H}_{\textrm{B-U}}}_{\textrm{BS-to-user link}} \nonumber\\&+ \sum_{n=1}^N\underbrace{\mathbf{H}_{\textrm{R-U}}^{(n)}}_{\textrm{RIS-to-user link}}\cdot\underbrace{\boldsymbol{\Theta}^{(n)}}_{\textrm{RIS steering}}\cdot\underbrace{\mathbf{H}_{\textrm{B-R}}^{(n)}}_{\textrm{BS-to-RIS link}}\,,
\end{align}
wherein $\mathbf{H}_{\textrm{B-U}} =  \left[\,\mathbf{h}_{\textrm{B-U},1}\; \mathbf{h}_{\textrm{B-U},2}\;\cdots\;\mathbf{h}_{\textrm{B-U},K}\,\right]^{\rm T} \in\mathbb{C}^{K\times L}$, with $\mathbf{h}_{\textrm{B-U},k} \in \mathbb{C}^{L\times 1}$ modeling the direct link between BS antennas and the $k$-th user equipment/terminal, $\mathbf{H}_{\textrm{R-U}}^{(n)} =  \left[\,\mathbf{h}_{\textrm{R-U},1}^{(n)}\; \mathbf{h}_{\textrm{R-U},2}^{(n)}\;\cdots\;\mathbf{h}_{\textrm{R-U},K}^{(n)}\,\right]^{\rm T} \in\mathbb{C}^{K\times Q^{(n)}}$, with $\mathbf{h}_{\textrm{R-U},k}^{(n)}[t] \in \mathbb{C}^{Q^{(n)}\times 1}$ modeling the direct link between the $n$-th RIS's reflective elements and the $k$-th user equipment/terminal,  $\boldsymbol{\Theta}^{(n)} = {\rm diag}\left\{\boldsymbol{\theta}^{(n)}\right\} \in\mathbb{C}^{Q^{(n)}\times Q^{(n)}}$ comprises the reflection coefficients of the $n$-th RIS, and $\mathbf{H}_{\textrm{B-R}}^{(n)}\in\mathbb{C}^{Q^{(n)}\times L}$ models the link between BS antennas and the $n$-th RIS's reflective elements. 

\subsection{Signal processing at the transmitters}\label{sub:exist-RIS-PC}

The role of the signal processing at transmitters is to produce message-related signals which, after passing though the communication channel, can be reliably reconstructed by sufficiently empowered receivers. When multiple transmit antennas are available, a basic processing scheme to increase spectral efficiency is beamforming, so that the transmit signals are steered toward specific directions. Alternatively, when channel-state information (CSI) is available, constructive overlap of the signals propagating through different paths can be induced at the receivers via proper precoding at the transmitter.

In MU scenario, where communication-related resources (e.g. physical channel, available transmit power, etc.) are shared among different users, the signal processing chain at the transmitter side is responsible for dealing with interference as well as possibly implementing resource-allocation strategies.  Indeed, considering the efficient usage of the available resources, there is a natural tendency for MU communication systems to work in an interference-limited regime. Traditionally, interference is considered to be a performance-impairing factor in wireless communications and, for this reason, precoding techniques (key components of the signal processing chain at the transmitter of downlink systems) were originally developed to mitigate or cancel out multiuser interference (MUI) using CSI. Nonetheless, information theory shows that known interference does not affect the underlying channel capacity~\cite{costa1983dpc}. Thus, by trying to eliminate the interference, the overall system performance might be reduced, if traditional channel-level precoding is applied. This type of precoding refers to the data-independent precoding strategy, which is solely based on the available CSI. In fact, the information-theoretic approach in~\cite{costa1983dpc} shows that the optimal transmitter adapts its signal to the interference instead of attempting to cancel it  out. Following this approach, symbol-level precoding (SLP) accounts for the underlying MUI by shaping the transmitted waveforms so as to induce constructive interference (CI) at each user. SLP is a non-linear technique that employs CSI along with users' data to form the precoder~\cite{8359237,9035662}.

By collecting the symbols to be transmitted for each one of the $K$ users during a symbol interval in the vector $\mathbf{s} \in\mathbb{C}^{K\times 1}$, one can obtain the transmitted signal vector via precoding as follows:
\begin{align}\label{eq:xDL}
    \mathbf{x} = \mathbf{P}\cdot\mathbf{s}\,,
\end{align}
in which $\mathbf{P} \in \mathbb{C}^{L\times K}$ is the so-called precoding matrix. For channel-level precoders, $\mathbf{P}$ does not depend on the particular symbols to be transmitted, whereas for symbol-level precoders $\mathbf{P}$ is determined in function of $\mathbf{s}$. The possible performance advantages from adopting an SLP scheme comes at the price of the computational complexity. Both types of precoders require CSI information for designing $\mathbf{P}$. 

Some channel-level precoding techniques have already been proposed~\cite{GuoTWC2020,DiJSAC2020} for MU multiple-input single-output (MISO) RIS-assisted downlink systems. These works use efficient algorithms to maximize the sum-rate of all users by jointly optimizing the BS precoder, by designing $\mathbf{P} = \left[\,\mathbf{p}_{1}\; \mathbf{p}_{2}\;\cdots\;\mathbf{p}_{K}\,\right]$, and the RIS passive beamformer, by designing $\boldsymbol{\theta}^{(1)}$. Both works assume that only $N = 1$ RIS is available and that the BS has a perfect CSI knowledge. More specifically, by defining the signal-to-interference-plus-noise ratio (SINR) of the $k$-th user as
\begin{align}
\label{eq:SNR}
    \gamma_k \!=\! \frac{\left\vert\left(\mathbf{h}_{\textrm{B-U},k}^{\rm T} \!+\!  \left(\mathbf{h}_{\textrm{R-U},k}^{(1)}\right)^{\rm T}\!\!\boldsymbol{\Theta}^{(1)}\mathbf{H}_{\textrm{B-R}}^{(1)} \!\right)\mathbf{p}_{k}\right\vert^2}{\sum\limits_{\substack{k'=1\\k'\neq k}}^K\!\left\vert\left(\mathbf{h}_{\textrm{B-U},k}^{\rm T} \!+\!  \left(\mathbf{h}_{\textrm{R-U},k}^{(1)}\right)^{\rm T}\!\!\boldsymbol{\Theta}^{(1)}\mathbf{H}_{\textrm{B-R}}^{(1)} \!\right)\!\mathbf{p}_{k'}\right\vert^2 \!+\! \sigma_{w,k}^2}
\end{align}
and 
given some weights $\omega_k > 0$, with $k\in\{1, 2, \dots, K\}$, representing relative priorities across different users, these works initially targeted to solve the problem\footnote{So far, it is typically assumed that the optimization is carried out centrally at the BS. Nevertheless, future system configurations may enable a distributed operation of RIS as well. This option will be addressed in Section \ref{sec:distr}. In addition, the issues related to the design of control signaling from the BS to the RISs are discussed in Section \ref{sec:control}.}
\begin{subequations}\label{eq:optProblemCLP}
\begin{alignat}{2}
&\!\mathop{\text{maximize}}\limits_{\mathbf{P},\boldsymbol{\theta}^{(1)}}        &\qquad& \sum_{k=1}^K \omega_k \log_2\left(1 + \gamma_k\right)\label{eq:optProb-CLP}\\
&\text{subject to} &      & \theta_{q}^{(1)} \in {\cal F},\; \forall q\label{eq:constraint1-CLP}\\
&                  &      & \sum_{k=1}^K\Vert\mathbf{p}_{k}\Vert_2^2 \leq \bar{P},\label{eq:constraint2-CLP}
\end{alignat}
\end{subequations}
in which $\bar{P}$ models a maximum transmit power constraint that must be satisfied by the BS precoder, whereas ${\cal F}$ denotes the feasible set of reflection coefficients, which models the additional constraint the RIS passive beamformer has to satisfy. The feasibility set can be defined in three different ways, namely: (i) general reflection ${\cal F} = \left\{\theta \in \mathbb{C} \,\left|\right.\, |\theta| \leq 1\right\}= \left\{\eta{\rm e}^{{\rm j}\phi}\,\left|\right.\, (\eta,\phi) \in [0,1]\times [0,2\pi)\right\}$, (ii) continuous-phase shifter  ${\cal F} = \left\{\theta \in \mathbb{C} \,\left|\right.\, |\theta| = 1\right\} = \left\{{\rm e}^{{\rm j}\phi}\,\left|\right.\, \phi \in [0,2\pi]\right\}$, and (iii) discrete-phase shifter  ${\cal F} =  \left\{{\rm e}^{{\rm j}\phi}\,\left|\right.\, \phi \in \left\{0, \frac{2\pi}{\tau},\dots,\frac{2\pi(\tau-1)}{\tau}\right\}\right\}$, which models the practical case of finite-size RIS with a limited number, $\tau \in\mathbb{N}$, of discrete phase-shifts. Both works in~\cite{GuoTWC2020,DiJSAC2020} proposed convenient surrogates to the general problem in~\eqref{eq:optProblemCLP} so that they could find suboptimal solutions under different assumptions. Their results suggest that the RIS-assisted framework provides significant energy efficiency gains compared to conventional relay-assisted communications.  


Moreover, symbol-level precoding (SLP)  techniques working in an MU RIS-assisted scenario have already been proposed in~\cite{Liu2019,Liu2}. The work in~\cite{Liu2019} considers a scenario, where a RIS passively modulates a carrier produced by an RF signal generator; a passive beamforming method is then designed to minimize the maximum symbol error rate (SER) across the multi-antenna users to be served. On the other hand, the work in~\cite{Liu2} follows the traditional SLP strategy while optimizing the reflective elements on a channel-level basis. The authors approach the problem by considering an $M$-sized constellation for which they define the matrix $\mathbf{X} \in \mathbb{C}^{L\times M^K}$, whose columns collect all possible precoded signals obtained from~\eqref{eq:xDL}.
The targeted optimization problem is 
\begin{subequations}\label{eq:optProblemSLP}
\begin{alignat}{2}
&\!\mathop{\text{minimize}}\limits_{\mathbf{X},\boldsymbol{\theta}^{(1)}}        &\;\;& \Vert\mathbf{X}\Vert_{\rm F} \label{eq:optProb-SLP}\\
&\text{subject to} &      & \theta_{q}^{(1)} \in {\cal F},\; \forall q,\label{eq:constraint1-SLP}\\
&                  &      & \mathbf{\tilde{y}}_{k} \!=\! \mathbf{X}^{\rm T}\!\!\left(\!\mathbf{h}_{\textrm{B-U},k} \!+\!  \left(\!\mathbf{H}_{\textrm{B-R}}^{(1)}\!\right)^{\rm T}\!\!\!\boldsymbol{\Theta}^{(1)}\mathbf{h}_{\textrm{R-U},k}^{(1)} \!\right)\!, \forall k\label{eq:constraint2-SLP}\\
&                  &      & \left[\Re\left\{\tilde{y}_{k,i}{\rm e}^{-{\rm j} \angle s_{k,i}}\right\}-\sigma_{w,k}\sqrt{\gamma_k}\right]\tan\varphi \nonumber\\
&                  &      &-\left\vert\Im\left\{\tilde{y}_{k,i}{\rm e}^{-{\rm j} \angle s_{k,i}}\right\}\right\vert \geq 0, \; \forall k,i,\label{eq:constraint3-SLP}
\end{alignat}
\end{subequations}
in which $\cal F$ models a continuous-phase shifter, $\tilde{y}_{k,i} = \left[\mathbf{\tilde{y}}_{k}\right]_i$, $i \in\{1, 2, \dots, M^K\}$, denote all possible noise-free received signals at the $k$-th user equipment, $\varphi > 0$ is an angle that defines the region of constructive interference for the received PSK constellation, and the constraints in~\eqref{eq:constraint3-SLP} enforces CI. Thus, the reflective elements are optimized globally so that CI is attained, irrespective of the actual transmitted symbols ${\bf s}$. As in the channel-level solutions, the work in~\cite{Liu2} actually solves some convenient surrogates for the problem in~\eqref{eq:optProblemSLP}. Resource allocation is also tackled in~\cite{Liu2} by solving an alternative QoS balancing problem, aiming to maximize the minimum QoS for a given average transmit power budget.

\subsection{Signal processing at the receivers}
The shaping of the environment not only affects the design of the transmitting devices, but also the design of the receivers. In fact, controlled reflections from RIS can be used to reduce multipath and improve the signal quality, such that the complexity of the signal detection is generally reduced compared to the environments without RISs. 

In contrast to  conventional communication systems that employ relays equipped with sufficient signal processing capabilities, RISs are usually nearly passive and equipped with minimal on-board signal processing components. This makes channel estimation a key challenge for the proper deployment of smart  environments, given the massive number of links that need to be estimated as well as the cascaded structure of the dyadic backscatter channel, which is formed by the forward and backscatter links~\cite{Liang2019}. Therefore, novel algorithms and protocols should take into account the complexity of RISs and avoid heavy signal processing operations running on RISs.

Most of the prior works on RIS mainly focus on designing the reflection coefficients under the assumption of perfect CSI~\cite{Huang2018,Cui2019,Pan2019,xu_2020,wang2020}, which facilitates the derivation of system performance upper bounds,  but it is still unfeasible in practice. On the other hand, some works have recently proposed novel channel estimation methods for RIS-assisted single-user communication systems. For instance, a channel estimation approach is developed using least-squares estimator for a passive RIS-assisted system that serves a single-antenna energy harvesting user via a multi-antenna power beacon~\cite{Mishra2019}. In~\cite{jensen2019}, a channel estimation approach for a passive RIS-aided MISO communication system has been designed based on a minimum variance unbiased (MVU) estimator, where the RIS's elements follow an optimal series of activation patterns. Moreover,~\cite{Zheng2019} has investigated channel estimation for a RIS-enhanced orthogonal frequency-division multiplexing (OFDM) system for a single user served by the full reflection of the RIS at all time, i.e. all RIS's elements are switched ON with maximum reflection amplitude during both channel estimation and data transmission phases. In~\cite{nadeem2019intelligent}, a channel estimation protocol based on the minimum mean-squared error (MMSE) estimator has been proposed. 


While the aforementioned works aim to provide a sufficient channel estimation accuracy with the least complexity, some of the more recent works have proposed more complex estimator structures in order to better exploit the flexibility associated with RIS. A two-stage estimator has been introduced in~\cite{mirza2019channel} as a combination of the conventional MMSE-based MIMO channel estimator and the so-called bilinear adaptive vector approximate message-passing (BAdVAMP) scheme based on maximum-likelihood and MMSE metrics. More recently, a novel grouping and partitioning method for the hierarchical channel estimation has been proposed in~\cite{you2020channel}, where multiple RIS's reflective elements receive the same control signal and operate as one. 
Apart from the channel estimation, further signal propagation parameters need to be estimated as well, such as frequency and timing synchronization among transmitters and receivers. In this context, RISs can help to reduce the Doppler effect, as it has been shown in~\cite{basar2019reconfigurable}.



\section{Use cases for RIS in Smart Cities}
\label{sec:3}

In this section, we propose some potential deployment scenarios for RIS in Smart Cities, and discuss their corresponding implications for the signal propagation, highlighting advantages and opportunities. We make a basic distinction between stationary scenarios, where RISs are attached to immovable objects, and mobile scenarios otherwise.

\subsection{Stationary scenarios}
	RISs can be appropriately configured to steer the non-line-of-sight (non-LoS) links into low coverage areas, where LoS transmissions are impossible due to obstructions/shadowing. Similarly, RISs may be configured to redirect some undesirable signals to avoid interference with other concurrent communications, or creating destructive interference toward the malicious users. Additionally, the autonomous stationary low-power IoT devices (sensors and actuators) can benefit from RIS deployment for sustainable recharging through concentrating the energy toward them. Accordingly, several potential applications can be foreseen, where RISs can be deployed as part of the stationary environment of Smart Cities. Among the most relevant applications, we identify the following:
	\subsubsection{Smart homes} RISs can be deployed in the interior walls of homes with diverse sizes in order to enhance the local connectivity of numerous kinds of devices that rely on wireless connectivity for operation. Smart homes are an especially promising application of RIS due to the short distances between the devices, such that the RIS-reflected signal path is not much longer (and correspondingly not much more attenuated) compared to the direct path. Hence, the main purpose of RISs in this context is to enhance the spectral efficiency via constructive interference.
	\subsubsection{Smart buildings} The facades of large buildings may be coated with RISs to offer opportunities for the coverage enhancement and spectral efficiency increase. Correspondingly, the smart environment can provide a better integration of mobile objects including pedestrians and vehicles into the Smart City. Furthermore, smart buildings represent an interface between the indoor and outdoor entities, which should facilitate the access of private households to the public domains. Hence, the deployment of RISs in smart buildings should follow an adequate strategy that suits the purpose of the Smart City, i.e. starting with the buildings in areas with many potential smart home customers and developed outdoor infrastructure.
	 \subsubsection{Smart factories} Smart factories have been envisioned with the concept Industry 4.0, which relies on massive machine-type communication. Just like in traditional IoT, RISs can help extend the coverage, thus avoiding granular clustering of the network. However, the application of RIS in smart factories is even more beneficial. In fact, the presence of large metallic objects in some industrial premises makes the wireless propagation environment very destructive for communications. However, RIS can establish a suitable approach for fine-tuning the signal reflections to find a path around the obstacles and thus increase the coverage even in very harsh environments.
	 \subsubsection{Smart hospitals} While a large number of sensors might be deployed in future hospitals, this massive connectivity may not be the main purpose of RIS in this application. Hospitals are more sensitive to electromagnetic radiations so that the radio intensity has to be sufficiently low to comply with very stringent regulations. In such scenarios, RIS can play a crucial role to control the harmful radiations without compromising communication quality by redirecting the signals away from sensitive areas.
	 \subsubsection{Smart billboards} Billboards in Smart Cities may be considered as an interesting option for RIS deployment of different sizes and heights, which provide the connectivity to a large number of users simultaneously whether they are indoors or outdoors.
	
	In the context of deploying stationary RIS, several advantages can be obtained in comparison with the mobile scenarios, e.g., more accurate CSI, predicting the signal propagation through stationary environments, and efficient optimization to the transmit beamforming and RIS phase shifts. Moreover, the quality of the end-to-end channels can be characterized and optimized through RIS physical parameters such as size, position, and orientation. In order to deliver the best communication performance, RIS should be fully aware of the complex and non-stationary surrounding environments where they are deployed, and thus, the spatial distribution of RISs has to be optimally configured for the best suitable functioning. However, this necessitates extra intelligent resources on the network to tackle the deployment complexity, real-time planning/optimization, and dynamic control.

\subsection{Mobility scenarios}
The mobility in RIS-assisted communication networks has typically been covered with respect to UAVs. So far, most of the relevant works  address this system aspect in terms of tracking capabilities using the passive beamforming of stationary deployed RISs. Recently, the mobility of RISs themselves has been addressed, where RISs have been attached to a UAV in order to combine the degrees of freedom associated with the UAV and with the RISs, respectively~\cite{lu2020enabling}. In general, the potential of RIS-enhanced mobile infrastructure is very high due to a higher chance for an LoS connectivity. However, the corresponding system complexity might also be very high. In fact, permanent re-synchronization of transmissions and channel estimations are needed in order to be able to benefit from the spatial diversity and high signal quality. These and other system aspects may depend on the type of RIS mobility. 
In the following, we discuss some of the possible mobility profiles that can appear in the context of Smart Cities.
\subsubsection{Stochastic mobility}
Stochastic mobility is associated with the random motion of mobile objects, especially end-users and uncontrolled UAVs. In this context, we may assume a certain probability for each direction of motion, which leads to the variation of the channel gain. Hence, a probability density for the channel evolution can be obtained. As a result, robust optimization seems promising for the design of the passive beamforming.
\subsubsection{Steerable mobility}
Steerable mobility is envisioned in the context of  UAVs. In fact, mobile RISs with a known and adjustable trajectory may dramatically improve the connectivity, since the UAV would act as a passive mobile relay. The benefit of this approach is almost no power consumption as well as very high network throughput. The latter stems from the fact that the relay does not need to receive the signals but solely reflect them. Correspondingly, it is unlikely to become a bottleneck for the network performance. In addition, RIS can be flown to a better position with a bigger impact for the signal quality, which is very beneficial as well. Correspondingly, this steerable mobility can be included in the design of the communication system as partially based on channel prediction and partially based on control loop automation.
\subsubsection{Predictable mobility}
Predictable mobility refers to the mobility of the infrastructure, which would follow a predefined route according to some known rules, however, without the possibility to adjust or control the trajectory. Within the scope of Smart City, this mobility corresponds to the public transportation, such as buses, trams, etc. This large mobile infrastructure can be equipped with RIS and act as a relay as well. Hence, the main functionality of the mobile RIS would be preserved. However, unlike steerable RISs, public transportation is not controlled by the communication system, since it has a completely different primary purpose. On the other hand, the mobility is predictable, since all buses or trams of the same lines follow the same route. Correspondingly, the evolution of the communication channels and reflection coefficients can be predicted. Furthermore, the travel route can be described as a sequence of states, such that the channel variations can be predicted via specifically designed Markov chains or using machine learning methods.
\subsubsection{Hybrid mobility}
Hybrid mobility describes mobile parts of the Smart City infrastructure, which do not fall in any of the previous mobility category. As an example, private vehicles equipped with RIS may partially follow the route of the public transportation, especially buses. Partially, these vehicles would resemble a stochastic mobility due to a large number of possible decisions and correspondingly states of a Markov process. Of course, the balance between predictability and randomness in these cases depends on the actual street scenario, such that in many cases there is no legal option for changing the path. However, this type of mobility may still differ from the well-defined predictable mobility due to varying sizes of RISs and speed of their motion.

The above scenarios for the application of RISs in Smart Cities introduce additional restrictions on the design of RIS-assisted networks. Hence, each research challenge discussed in Section~\ref{sec:4} can be addressed independently for each of the aforementioned scenarios. For each combination of a challenge and a use case, the respective optimization problems may differ according to the specific restrictions imposed by the use case, which may dramatically affect the way such problems are solved. As an example, channel estimation in scenarios with predictable mobility does not follow the simple MMSE solution explained earlier, but may need to be incorporated into a specifically designed Markov chain. In the following, we discuss the existing and future research challenges, which can be tackled with respect to the aforementioned scenarios. 

\section{Existing Research Challenges}
\label{sec:4}
In this section, we describe some of the urgent research challenges, which need to be addressed in the near future in order to benefit from the promising functionalities of RISs in Smart Cities. We also provide the reader with some concrete examples on how to approach those challenges. 
\subsection{Pilot decontamination}
As explained previously, accurate channel estimation is one of the essential prerequisites for the correct operation of RIS. In this context, pilot contamination is one of the most performance detrimental issues in large antenna-array systems, e.g., massive MIMO systems~\cite{Marzetta2010,pilot2019}. Due to a large number of reflective elements, which resemble passive antennas, this problem occurs in RIS-assisted networks as well. Since RISs will be more densely distributed than BSs in Smart Cities, pilot contamination in RIS-assisted systems will be more complicated and different than in the massive MIMO systems. Therefore, more comprehensive studies should be conducted for pilot decontamination in RIS-assisted systems because the existing methods developed for massive MIMO applications may not be effective~\cite{Jung2019}. 
For instance, pilot decontamination can be reached via efficient pilot assignment schemes with the objective of maximizing the minimum average signal to interference ratio among the served users. In our studied system model, as the number of BS antennas grows large, the optimization problem can be formulated as
\begin{subequations}\label{eq:pilot_decontamination}
\begin{alignat}{2}
	&\ \!\!\!\!\!\hspace*{1mm}\mathop{\text{maximize}}\limits_{\mathcal{P}_s} &&\hspace*{-1mm}\mathop{\text{min}}\limits_{\forall k} \left\lbrace  \lim_{L \to \infty} \frac{\left|  \left(\mathbf{h}_{\textrm{R-U},k}^{(n)}\right)^{\rm \!\! T} \boldsymbol{\Theta}^{(n)}\mathbf{H}_{\textrm{B-R}}^{(n)} \mathbf{s}_k\right|^2}{\!\!\sum\limits_{\substack {n^\prime=1,n^\prime \neq n}}^N 
		\left| \left(\mathbf{h}_{\textrm{R-U},k}^{(n^\prime)}\right)^{\rm \!\! T} \boldsymbol{\Theta}^{(n^\prime)}\mathbf{H}_{\textrm{B-R}}^{(n^\prime)}\mathbf{s}_k\right|^2}  \right\rbrace \label{eq:pilot_decontamination_C1}\\
	&\text{subject to:} & & \;\; \left| \theta_{q}^{(n)} \right| =1,\; \forall q,  \label{eq:pilot_decontamination_C2}
\end{alignat}
\end{subequations}

\noindent
where $\mathcal{P}_s$ denotes the set of pilot sequences $\mathbf{s}_k$, and here the direct link between the BS and the $k$-th user was assumed to be blocked.
Additionally, obtaining the full channel estimates for all reflective elements in RIS is a prohibitively difficult process given the restricted channel training time. Hence, an optimization of the length and the number of pilot sequences that are required for a precise CSI with a moderate training overhead is an important research direction.  
	
	Furthermore, in RIS-enhanced propagation environments, channel characteristics substantially differ from the conventional channel models. 
	Specifically, in a Smart City scenario, some of the network nodes including RISs may exhibit unique channel impairments, which should be taken into account when developing the respective channel models. More importantly, there are some challenges related to the signal propagation that are overlooked in the existing literature, e.g., the presence of reflective arrays of arbitrary sizes that may increase the probability of polarization mismatches over the large arrays. Moreover, near- and far-field behavior as well as the effective areas of the RIS's elements have been  disregarded in RIS related studies so far. Future research may consider developing a channel-modeling framework that takes into consideration the various signal propagation characteristics in mobile RIS scenarios, and then devising channel estimators for dynamic channel conditions to cover all possible system setups that would benefit from the RIS concept. 
	
	In order to address channel prediction and to enhance  accuracy, the entire channel training phase may need to incorporate all the aforementioned distinctive channel characteristics of RIS-assisted networks. Specifically, novel training schemes need to be developed, which account for the spatial and temporal channel variations in stationary and mobile scenarios. 
	Particularly, a distinction among steerable, predictable, and stochastic mobility of devices has to be cautiously addressed. As an example, the characterization of dynamic UAV channels has to be taken into account in developing channel estimation strategies with their special propagation characteristics, such as airframe shadowing caused by the UAV's structural design and rotating capabilities.

\subsection{Control signaling}
\label{sec:control}
One of the main challenges, which has been mostly neglected so far, is the remote control of the RIS's elements. In fact, it has been simply assumed that the optimized phase shifts can be uploaded into the memory of the RIS's microcontroller. This approach, however, requires a well synchronized, and possibly dedicated, control link between BS and RIS. While an almost perfect synchronization can be assumed for stationary scenarios, it may become a real challenge in case of RIS mobility due to the time-varying channels and fading effects. 

Also, in presence of multiple RISs, the co-channel interference from multiple control signals may lead to imperfect adaptation of RISs and thus a quick performance degradation. In order to resolve this issue, orthogonal channels might be used, thus reducing the overall system capacity, or otherwise very strong forward error correction (FEC)  codes and adaptive frequency plan might be used. However, FEC codes with low coding rate impose a high  latency, which is not desirable.

Interestingly, the update frequency for the RIS's elements has an impact on the coherence time of the communication channels, such that the coherence time is no longer solely related to the physical processes within the environment, e.g. stationary reflection or slowly moving obstacles~\cite{rappaport}. Correspondingly, the faster the adaptation of RIS, the smaller is the coherence time of the channels. On the other hand, the update frequency depends on the time-varying processes in the network. This leads to a trade-off between the signal quality and the accuracy of the known system parameters.

Another problem for the design of the control signaling is the amount of data to be transmitted to RIS. Depending on the size of the surface, each RIS may contain hundreds of reflective elements, which need to be adjusted. Correspondingly, the power consumption associated with the control signals may not be negligible as typically assumed. Furthermore, in order to achieve a timely update of all RIS's reflective elements, the signal bandwidth needs to be sufficiently large. One possible solution to both high-power and bandwidth consumption is to reduce the quantization resolution, such that the amount of data to be transmitted is reduced at the cost of lower accuracy. An alternative solution would be to cluster the panels of each RIS, such that the updates may be executed per cluster instead of per panel. In this context, the optimal number of clusters, the shape of each cluster, and the update protocol need to be designed in accordance with the network requirements. 
Thus, assuming a predefined maximum number of clusters $R^{(n)} < Q^{(n)}$, for $n\in\{1, \dots, N\}$, then a clustering of the set of indexes  $\{1, \dots, Q^{(n)}\}$ of the RIS's reflective elements can be denoted by the sets ${\cal Q}^{(n)}_1, \dots, {\cal Q}^{(n)}_{R^{(n)}}$, which satisfy two conditions, viz.: (i) ${\cal Q}^{(n)}_1\cup \cdots\cup {\cal Q}^{(n)}_{R^{(n)}} = \{1, \dots, Q^{(n)}\}$ and (ii) ${\cal Q}^{(n)}_r\cap {\cal Q}^{(n)}_s = \emptyset$ for $r \neq s$. In this case, all  reflection coefficients of the $n$-th RIS is parameterized by $\boldsymbol{\tilde{\theta}}^{(n)} \in \mathbb{C}^{R^{(n)}\times 1}$. Taking these definitions into account, then a possible problem formulation for the clustering of the panels is similar to~\eqref{eq:optProblemCLP}, as follows:
\begin{subequations}\label{eq:optCluster}
\begin{alignat}{2}
&\!\mathop{\text{maximize}}\limits_{\mathop{\mathbf{P}}\limits_{\mathop{\boldsymbol{\tilde{\theta}}^{(1)}, \dots, \boldsymbol{\tilde{\theta}}^{(N)}}\limits_{{\cal Q}^{(n)}_1, \dots, {\cal Q}^{(n)}_{R^{(n)}}, \forall n}}}        & & \sum_{k=1}^K \omega_k \log_2\left(1 + \gamma_k\right)\label{eq:optCluster2}\\
&\text{subject to}&      & \sum_{k=1}^K\Vert\mathbf{p}_{k}\Vert_2^2 \leq \bar{P},\label{eq:optCluster3}\\
&                  &      & \theta_{q}^{(n)}=\tilde{\theta}^{(n)}_r \in {\cal F},\; \forall q\in{\cal Q}^{(n)}_r,\:\forall r,\:\forall n.\label{eq:optCluster4}
\end{alignat}
\end{subequations}
In this context, the number of clusters $R^{(n)}$ is subject to a distinct investigation. Obviously, the clustering strategy affects the precoding vector and power consumption.

\subsection{Precoding for large multiuser systems}

When it comes to precoding, the existing  techniques assume a simplified scenario with only one static RIS, which might not be suitable to some use cases described in Section~\ref{sec:3}. In contrast, multiple RISs can substantially increase the spatial diversity thus making precoding even more promising. For instance, one might target solving the optimization problem 
\begin{subequations}\label{eq:optProblemSLP-new}
\begin{alignat}{2}
&\!\mathop{\text{minimize}}\limits_{\mathop{\mathbf{x}}\limits_{\boldsymbol{\theta}^{(1)},\dots,\boldsymbol{\theta}^{(N)}}}        &\;\;& \Vert\mathbf{x}\Vert_2 \label{eq:optProb-SLP-new}\\
&\text{subject to} &      & \theta_{q}^{(n)} \in {\cal F},\; \forall n,q,\label{eq:constraint1-SLP-new}\\
&                  &      &  \left[\mathbf{H}\left(\boldsymbol{\theta}^{(1)},\ldots,\boldsymbol{\theta}^{(N)}\right)\cdot\mathbf{x}\right]_k \in \mathscr{C}_k, \; \forall k, \label{eq:constraint2-SLP-new}
\end{alignat}
\end{subequations}
in which $\cal F$ models general reflections, or continuous-phase shifters, or even discrete-phase shifters (see Section~\ref{sub:exist-RIS-PC}), 
$\mathbf{H}\left(\boldsymbol{\theta}^{(1)},\ldots,\boldsymbol{\theta}^{(N)}\right)$ is defined in~\eqref{eq:HDL}, and $\mathscr{C}_k \subset \mathbb{C}$ defines the CI convex region in which the noise-free signal of the $k$-th user must lie. In addition to being dependent on the parameters $\sigma_{w,k}$ and $\gamma_k$, the CI regions $\mathscr{C}_k$ are constellation-dependent, and that is why we have omitted its analytical definition here. 

One should note that, although the dependency of the RIS's reflection coefficients on the symbol slot $t$ is not explicitly denoted, all optimization variables actually depend on $t$. Thus, solving the problem in~\eqref{eq:optProblemSLP-new} constitutes a real challenge in the practical scenarios wherein CSI knowledge is not always readily available/reliable and the configuration parameters of RIS cannot be updated in a symbol-level basis.    

Therefore, it is part of the existing research challenges to conceive meaningful surrogates for the problem in~\eqref{eq:optProblemSLP-new} to enhance both spectral and energy efficiencies using multiple RISs as (possibly mobile) relays. For this, optimization-based iterative methods as well as data-driven solutions might be considered.

In addition, resource allocation (e.g. power, subcarriers, etc.) must also be considered for RIS-assisted environments to further increase the energy efficiency of the system. More specifically,
\begin{itemize}
    \item power allocation techniques,
    \item user scheduling with the focus on  unicasting/multicasting schemes,
    \item carrier assignment in multicarrier communications for frequency-selective channels
\end{itemize}
are examples of problems to be addressed in the near-future for the envisioned use cases.

\subsection{Distributed operation}
\label{sec:distr}
Smart environments can be realized by deploying RISs in centralized, distributed, or hybrid manners. In centralized implementations, a central entity controls the communication-related activities of the other network entities. In this implementation, RISs are completely passive, since they only need to receive their configuration parameters from the central controller, and reflect the impinging signals accordingly. The central controller therefore needs to have reliable processing and communication capabilities to obtain CSI via some suitable protocols and, most importantly, without relying on RIS's own processing/communication capabilities. On the other hand, in distributed implementations of RISs, some additional sensing and processing functionalities are required for the surfaces to be able to autonomously optimize themselves based on the environmental state information and network configuration. This can be realized through boosting RIS's capabilities by embedding low-power sensors as well as communication and digital processors. Thus, the inherent signaling overhead of a centralized solution can be traded off by an increased computational burden over the RIS with the corresponding higher deployment costs.  In-between the centralized and distributed implementations, a hybrid implementation may employ RISs with different capabilities to trade off the  implementation complexity and power consumption of RISs in the distributed implementation with the higher channel estimation and signaling overhead that are necessary to feed the central controller in centralized implementations.

Let us consider a concrete example of distributed operation from a precoding perspective. Assume that, at the $t$-th symbol slot, a reasonable estimate of the BS precoding matrix $\mathbf{P}^{(n)}[t-1]$ is available at the $n$-th RIS along with CSI knowledge of both BS-to-RIS (i.e. $\mathbf{H}_{\textrm{B-R}}^{(n)}[t]$) and RIS-to-user (i.e. $\mathbf{H}_{\textrm{R-U}}^{(n)}[t]$)  links; then it is possible to obtain a local soft estimate $\mathbf{\hat{s}}^{(n)}[t] = \left(\mathbf{H}_{\textrm{B-R}}^{(n)}[t]\mathbf{P}^{(n)}[t-1]\right)^\dagger\mathbf{y}^{(n)}[t]$ of the transmitted symbols $\mathbf{s}[t]$, where $\mathbf{y}^{(n)}[t]$ is the RIS's received signal. Thus, for the $n$-th RIS, the following optimization problem can be targeted: 
\begin{subequations}\label{eq:optProblemSLP-dist}
\begin{alignat}{2}
&\!\mathop{\text{minimize}}\limits_{\mathbf{P}^{(n)}[t],\boldsymbol{\theta}^{(n)}[t]}        & & \Vert\mathbf{P}^{(n)}[t]\mathbf{\hat{s}}^{(n)}[t]\Vert_2 \label{eq:optProb-SLP-dist}\\
&\text{subject to} &      & \!\!\!\!\theta_{q}^{(n)}[t] \in {\cal F},\; \forall q,\label{eq:constraint1-SLP-dist}\\
&                  &      &  \!\!\!\!\!\!\!\!\!\!\left(\mathbf{h}_{\textrm{R-U},k}^{(n)}[t]\!\right)^{\rm T}\!\!\!\boldsymbol{\Theta}^{(n)}[t]\mathbf{H}_{\textrm{B-R}}^{(n)}[t]\mathbf{P}^{(n)}[t]\mathbf{\hat{s}}^{(n)}[t] \!\in\! \mathscr{C}_k^{(n)}\![t],  \forall k, \label{eq:constraint2-SLP-dist}
\end{alignat}
\end{subequations}
where $\mathscr{C}_k^{(n)}[t]\subset\mathbb{C}$ is the corresponding CI region parameterized by a given SINR $\gamma_k^{(n)}$,  satisfying $\sum_n \sqrt{\gamma_k^{(n)}} = \beta\sqrt{\gamma_k}$, for a  predefined parameter $\beta \in (0,1]$. Once all $N$ RISs solve this problem, they can self-configure according to the optimal $\boldsymbol{\theta}^{(n)}[t]$. Besides, at the symbol slot $t+1$, the $n$-th RIS can use as precoding matrix either: the optimal $\mathbf{P}^{(n)}[t]$ calculated locally at slot $t$, or a convenient combination of its neighboring (locally calculated) RIS's precoding matrices --- the neighboohood is formed so that several RISs jointly cooperate to enhance performance through dedicated signaling channels --- with its own locally calculated precoding matrix, or finally the actual BS  precoding matrix $\mathbf{P}[t+1]$,  which might be broadcasted from time to time, following a determined protocol that prevents RISs from getting accumulated performance degradation.   

Again, the problem in~\eqref{eq:optProblemSLP-dist} might be approached through convenient surrogates that best suit the available mathematical/computational tools. One might also consider formulating different problems using  data-driven approaches relying on (distributed) least-mean squares (LMS) and recursive-least squares (RLS) adaptive filtering algorithms for online beamforming, or using more recent artificial intelligence tools, like federated learning. Besides, the model itself might be generalized to account for multiple reflections on different RISs, as discussed in the next section.

\subsection{Hybrid relaying}

Cooperative communication is known to be beneficial for the throughput of wireless networks, since the overall path loss can be substantially reduced by placing a relay between source and destination. The signals received by the relay are typically amplified, (optionally) processed and then forwarded. Various configurations of relays are known in the literature~\cite{Liu3,Hossain}. Relaying methods differ mostly in the type of signal processing --- e.g. amplify-and-forward (AF), filter-and-forward (FF), and decode-and-forward (DF) relays --- and the availability of sufficient radio resources --- e.g. energy-constrained and buffer-aided relays. 

RISs are typically viewed as passive relays with the corresponding passive beamforming capabilities. However, as pointed out in Sections~\ref{sec:2} and~\ref{sec:distr}, signal processing at RIS is performed in order to make use of the control signaling or from the environment sensing. Hence, a very promising extension of the passive RIS-based relaying strategy is to combine passive and active relaying into a hybrid RIS-based relaying. The resulting optimization problem comprises the design of the RIS's elements and the characteristic parameters of the relay, e.g. amplification factor or buffer size. Assuming an amplification factor $\alpha > 0$, the active path can be described by the respective SINR
\begin{equation}
    \gamma_{k,{\rm act}}=\frac{\alpha^2|\mathbf{h}_{\textrm{R-U},{\rm act}, k}^{\rm T}\mathbf{H}_{\textrm{B-R}, {\rm act}}\mathbf{p}_{k}|^2}{\alpha^2\sum_{l\neq k}I_l+\alpha^2\|\mathbf{h}_{\textrm{R-U},{\rm act}, k}\|_2^2\sigma_z^2+\sigma_{w,k}^2},
\end{equation}
where $\mathbf{H}_{\textrm{B-R},{\rm act}}$ represents the channel matrix for the link connecting the BS antennas with the active antenna(s) of a RIS-relay. Moreover, $\mathbf{h}_{\textrm{R-U},{\rm act},k}$ represents the channel vector for the link between the active antenna(s) of a RIS-relay and the antenna of end-user $k$. The interference received by the relay is given by $I_l=|\mathbf{h}_{\textrm{R-U},{\rm act}, k}^{\rm T}\mathbf{H}_{\textrm{B-R}, {\rm act}}\mathbf{p}_{l}|^2$. Note that these matrices and vectors differ from the ones we assumed for the signal propagation via passive reflection from RIS. In addition, unlike passive relaying, active RIS-relay would amplify the received noise with the variance $\sigma_z^2$.

The SNR of the passive path is given in \eqref{eq:SNR}. Hence, a hybrid AF relaying problem can be formulated as follows:
\begin{subequations}\label{eq:optAFrelay}
\begin{alignat}{2}
&\!\mathop{\text{maximize}}\limits_{\mathbf{x},\alpha,\boldsymbol{\theta}^{(1)}}        &\qquad& \sum_{k=1}^K \omega_k \log_2\left(1 + \gamma_{k, \rm MRC}\right)\label{eq:optAFrelay2}\\
&\text{subject to}&      & \sum_{k=1}^K\Vert\mathbf{p}_{k}\Vert_2^2 \leq \bar{P},\label{eq:optAFrelay3}\\
&                  &      & \sum_{k=1}^K\alpha^2\left(\Vert\mathbf{H}_{\textrm{B-R}, {\rm act}}\mathbf{p}_{k}\Vert_2^2+\sigma_z^2\right) \leq \bar{P}_z,\label{eq:constraint1-CLP}
\end{alignat}
\end{subequations}
where $P_z$ represent the available power at the relay, while $\gamma_{k, \rm MRC}=\gamma_{k,{\rm act}}+\gamma_k$ denotes the maximum SNR obtained via maximum ratio combining (MRC) technique.
In this context, an important question to be addressed is regarding the preferred type of relaying and beamforming. As discussed earlier, the degree of freedom of RIS-assisted networks is much higher than for the same networks without RISs, which implies a very high complexity of the optimization. Hence, the dominant system aspects need to be identified in order to reduce the complexity whenever possible. This can be done by assessing the dependency of the system performance on passive and active signal paths and by evaluating the respective contributions to the resulting signal quality and data rate. More specifically, data-driven methods can be employed in order to determine the suitable scenarios, for which active, passive, or both relaying types are more beneficial.

\subsection{Security and Privacy control}
As mentioned earlier, privacy and security are of utmost importance in a Smart City. In order to enhance these system aspects, physical layer security (PLS) techniques can be employed. PLS is a popular research field, which has been used in various applications in order to enhance the secrecy and data protection against eavesdropping. Typically, artificial noise is transmitted in order to reduce the signal quality for the eavesdropper~\cite{mukherjee2014principles}.

Since RISs are responsible for the improvement of the signal quality via passive relaying, the same approach can be used to reduce the signal quality for the eavesdropper. However, passive RIS cannot emit artificial noise. Hence, other tactics are needed in order to reduce the signal quality at the target location. 

One of the promising methods is to apply passive beamforming of RIS in order to avoid signal reflections in the direction of a possible eavesdropper~\cite{yu2019robust}.

An alternative approach is to partially redirect some of the data streams in order to increase the amount of interference at the eavesdropper. Assume a single eavesdropper with $N_{\rm Eve}$ antennas and two legitimate users with respective channel $\textbf{H}_{1}\left(\boldsymbol{\theta}^{(1)},\ldots,\boldsymbol{\theta}^{(N)}\right)$ and
$\textbf{H}_{2}\left(\boldsymbol{\theta}^{(1)},\ldots,\boldsymbol{\theta}^{(N)}\right)$. In addition, assume that the CSI for the link between BS and the eavesdropper (that is interested only in the data of the first legitimate user) is perfectly known\footnote{Of course, the exact position of the eavesdropper is not available in practice, such that robust optimization is usually preferred. However, we assume perfect CSI for the eavesdropper in order to illustrate the proposed approach.} and described by a channel matrix $\textbf{H}_{\rm Eve}\left(\boldsymbol{\theta}^{(1)},\ldots,\boldsymbol{\theta}^{(N)}\right)$. We obtain the covariance of the received useful signal and the covariance of the received interference at the eavesdropper as 
\begin{eqnarray}
\hspace*{-5mm}\textbf{D}\hspace*{-3mm}&=&\hspace*{-3mm}\textbf{H}_{\rm Eve}\left(\boldsymbol{\theta}^{(1)},\ldots,\boldsymbol{\theta}^{(N)}\right)\textbf{p}_1\textbf{p}_1^{\rm H}\textbf{H}^{\rm H}_{\rm  Eve}\left(\boldsymbol{\theta}^{(1)},\ldots,\boldsymbol{\theta}^{(N)}\right)\hspace*{-1mm},\\
\hspace*{-5mm}\textbf{K}\hspace*{-3mm}&=&\hspace*{-3mm}\textbf{H}_{\rm Eve}\left(\boldsymbol{\theta}^{(1)},\ldots,\boldsymbol{\theta}^{(N)}\right)\textbf{p}_2\textbf{p}_2^{\rm H}\textbf{H}^{\rm H}_{\rm  Eve}\left(\boldsymbol{\theta}^{(1)},\ldots,\boldsymbol{\theta}^{(N)}\right)\hspace*{-1mm},
\end{eqnarray}
respectively. Accordingly, the channel capacity between the BS and the potential eavesdropper is given by
\begin{equation}
    C_{\rm Eve}=\log_2\det\left(\mathbf{I}_{N_{\rm Eve}}+\left(\sigma^2_{\mathrm{Eve}}\textbf{I}_{N_{\mathrm{Eve}}}+\textbf{K}\right)^{-1}\textbf{D}\right),
\end{equation}
in which $\sigma^2_{\mathrm{Eve}}$ denotes the power of the additive noise at the eavesdropper. 
The channel capacities for the legitimate links are given by $C_{1}=\log_2(1+\gamma_1)$ and $C_2=\log_2(1+\gamma_2)$, respectively. Hence, for the secrecy rate, we obtain $\left[C_1-C_{\rm Eve}\right]^+$. The optimization problem, which aims at increasing the secrecy of the first legitimate user while satisfying the demand $C_{\mathrm{demand}}$ of the second user is given by
\begin{subequations}\label{eq:optProblemSLP-dist}
\begin{alignat}{2}
&\!\mathop{\text{maximize}}\limits_{\mathop{\textbf{p}_1, \textbf{p}_2}\limits_{\boldsymbol{\theta}^{(1)},\dots,\boldsymbol{\theta}^{(N)}}}       & \quad& \left[C_1-C_{\rm Eve}\right]^+ \label{eq:optProb_PLS}\\
&\text{subject to} &      & C_2\geq C_{\mathrm{demand}},\label{eq:constraint1-PLS}\\
&                  &      & \theta_{q}^{(n)} \in {\cal F},\; \forall n,q,\label{eq:constraint2-PLS}\\
&                  &      & \Vert\mathbf{p}_{1}\Vert_2^2+\Vert\mathbf{p}_{2}\Vert_2^2 \leq \bar{P}\label{eq:constraint3-PLS}.
\end{alignat}
\end{subequations}
Of course, the effectiveness of this approach depends on the number of data streams, which are permitted to be utilized as interference. In the context of Smart City, the choice of the data streams to become interference needs to be made in accordance with the network management. As an example, open access services are well suited, since a high level of protection is not required for them.

\section{Future Research Challenges and Opportunities}
\label{sec:5}
After the initial deployment of RISs in Smart Cities, the functionalities and operation principles of RIS may go through multiple updates as it is common for communication systems. Hence, we can anticipate more advanced schemes and system configuration to become trendy research fields in the more distant future. In the following, we motivate some of these research directions.

\subsection{Mobile multihop relaying}
Multihop relaying (MHR) is a typical generalization of the basic relaying concept with multiple relays between source and destination. This concept 
has become a promising research direction with the introduction of wireless sensor and ad hoc networks~\cite{shorey2006mobile}. However, the main drawback of this generalization lies in the system design complexity, which increases with each relay.

In addition, the mobility of some relays represents an additional degree of freedom, which can be explored in the respective scenarios. Interestingly, the mobility of the nodes/relays still contributes to the increase of the network coverage and throughput despite the stochastic motion of relays~\cite{xiao2010mobile}. However, one of the main disadvantages of this type of network is its fast changing topology. Recently, this research area has re-gained an increased attention with the introduction of UAVs. The degrees of freedom associated with the motion of a UAV can be exploited to enhance the signal quality. In particular, the changes in signal propagation due to the motion of nodes/users can be accounted for by designing the optimal trajectories for multiple UAVs, which leads to higher probability of LoS transmission in each hop, lower path loss and higher throughput. 


The existing works involving RISs and UAVs focus mainly on active beamforming using UAVs and passive beamforming using static RISs, which are combined in a joint optimization. A more challenging and potentially more beneficial architecture may involve mobile RIS-assisted multihop relays. Here, the idea would be to attach RISs to multiple UAVs, such that each mobile relay can perform hybrid relaying/beamforming, as explained earlier. Furthermore, RISs can be attached to other mobile objects according to the mobility scenarios discussed in Section~\ref{sec:3}. In this context, beamforming and resource allocation need to be combined with multihop relaying by taking into account the mobility profiles. 

In addition, the impact of the relay mobility on the system performance needs to be analyzed. Imperfect channel state information accumulated over a large number of reflections and hops can have a very harmful impact on the signal quality and accuracy of the optimization solution. Hence, the maximum number of mobile relays and the type of relaying needs to be carefully selected for each use case.

\subsection{Wireless RIS networks}
We distinguish between the operation of RIS in centralized vs. distributed manner, as explained in Section~\ref{sec:distr}. In the former case, RISs have a sole functionality of actuators, which execute the commands from the BS. Hence, no complicated signal processing at the RIS is required in this case. In the latter case, RIS is capable of making own decisions and can be viewed as part of a wireless sensor and actuator network (WSAN), a self-organized network (SON) or an ad hoc network (AHN), where the nodes adjust their operation to reach a common goal. In other words, a wireless RIS network can be established, where RIS devices represent a sub-system, which has the task of improving the communication environment for other Smart City components. Of course, this requires additional intelligence for the design of RIS, since the sensing and communication capabilities are needed in each network node~\cite{akyildiz2010wireless}. In this context, the design of the resulting WSANs is very challenging due to the high degree of freedom associated with RISs.

Among the various challenges in the context of networking with RISs, the previously mentioned hybrid mobile multihop relaying needs to be further extended to ensure the connectivity despite the complexity for the system optimization. It seems to be inevitable that suboptimal schemes with reduced complexity will be used for a sufficiently low latency and low power consumption at the relays.  

In the context of Smart City, the integration of such WSANs requires a careful cross-layer design using the existing methods of WSANs, SONs and AHNs, which should coordinate their transmissions with the services provided by the Smart City. Specifically, the higher OSI layers are important in order to guarantee the connectivity within such networks, which is essential for the accurate shaping of the communication environment.

Furthermore, the self-configuration methods associated with the SONs should be incorporated in order to facilitate distributed operation and independent functionality of the network. However, due to the high complexity of self-configuration methods and RIS-assisted networks in general, the respective design problem may not be tractable analytically. Hence, machine learning and artificial intelligence will be the most promising tools of system design for wireless RIS networks in future.

\subsection{Powering of RIS devices}
In scenarios where RISs are part of an autonomous wireless network, the flexibility for the deployment of RISs is based on the independence from the BS in terms of signaling and resources. Specifically, the nodes enhanced by RISs are expected to use rechargeable batteries whenever possible. Since the power consumption is known to be very low, we can anticipate the application of energy harvesting (EH) and even simultaneous wireless information and power transfer (SWIPT). Note that the application of EH and SWIPT for the charging of low-power devices in presence of RISs has already been addressed~\cite{wu2019weighted,zheng2020intelligent}, while the powering of RISs themselves is a novel research opportunity. Although the design of highly reflective surfaces coupled with energy harvesting seems to be very challenging, a combination of both in terms of a clever arrangement of the reflective elements is realistic. In this context, the optimization problem for the passive beamforming, relaying, etc. becomes even more complicated than discussed earlier, since parts of the signal might need to be reflected toward adjacent RISs for charging. A distinct advantage of this technique would be that parts of the electromagnetic radiation will be consumed by the surfaces, such that the unnecessary radiation is removed from the environment leading to a green environment and Green Smart City. This provides an additional motivation for the design of such low-power RIS-assisted networks.

\section{Conclusion}
\label{sec:6}
This paper brought up several practical and thought-provoking research directions to integrate RIS technology into wireless communication platforms of Smart Cities. First, the most promising scenarios and use cases for the deployment of RIS-assisted communication networks and services were introduced.  Based on these scenarios, several potential research challenges and future opportunities for the application of RISs in Smart Cities were proposed. Then, the key enabling designs of RIS-assisted networks were discussed, spanning a wide-ranging of problems and challenges that need to be tackled to push forward toward the digital transformation of Smart Cities. Specifically, pilot decontamination, precoding for multiple surfaces, and distributed operation are foreseen to require more immediate research efforts. Furthermore, new application scenarios of future RIS-assisted Smart Cities with novel features and properties were presented. Through this, the unprecedented research challenges that arise based on this integration were discussed. In short, this paper introduced several practical aspects of RIS-assisted communication networks that can potentially trigger more in-depth thinking and enrich the state-of-the-art of Smart City communications and smart wireless environments.
\bibliographystyle{IEEEtran}
\bibliography{IEEEabrv,Literature}
\end{document}